\documentclass[pra,twocolumn,showpacs,nofootinbib,letterpaper,superscriptaddress]{revtex4}

\usepackage{amsmath} % need for subequations
\usepackage{graphicx} % need for figures
\usepackage{verbatim} % useful for program listings
\usepackage{color} % use if color is used in text
\usepackage{subfigure} % use for side-by-side figures
\usepackage{hyperref} % use for hypertext links, including those to external documents and URLs

\begin{document}

\title{Absorption and emission of single attosecond light pulses in an autoionizing gaseous medium dressed by a time-delayed control field}
\date{\today}
\author{Wei-Chun Chu}
\affiliation{J. R. Macdonald Laboratory, Department of Physics, Kansas State University,
Manhattan, Kansas 66506, USA}
\author{C.~D.~Lin}
\affiliation{J. R. Macdonald Laboratory, Department of Physics, Kansas State University,
Manhattan, Kansas 66506, USA}
\affiliation{Department of Physics, National Taiwan University, Taipei 10617, Taiwan}

\begin{abstract}
An extreme ultraviolet (EUV) single attosecond pulse passing through a laser-dressed dense
gas is studied theoretically. The weak EUV pulse pumps the helium gas from the ground state
to the $2s2p(^1P)$ autoionizing state, which is coupled to the $2s^2(^1S)$ autoionizing state
by a femtosecond infrared laser with the intensity in the order of 10$^{12}$~W/cm$^2$.
The simulation shows how the transient absorption and emission of the EUV are modified by the
coupling laser. A simple analytical expression for the atomic response derived for
$\delta$-function pulses reveals the strong modification of the Fano
lineshape in the spectra, where these features are quite universal and remain valid for
realistic pulse conditions. We further account for the propagation of pulses in the medium
and show that the EUV signal at the atomic resonance can be enhanced in the gaseous medium by
more than 50\% for specifically adjusted laser parameters, and that this enhancement persists
as the EUV propagates in the gaseous medium. Our result demonstrates the high-level control of
nonlinear optical effects that are achievable with attosecond pulses.
\end{abstract}

\maketitle

%%%%%%%%%%%%%%%%%%%%%%%%%%%%%%%%%%%%%%%%%%%%%%%%%%%%%%%%%%%%%%%%%%%%%%%%%%%%%%%%%%%%%%%%%%%%%%%
\section{Introduction}\label{sec:intro}

The rapidly developing technologies of ultrashort light sources have gained wide applications
in the past 10 years~\cite{krausz, sansone}, most notably with the use of attosecond pulses in
time-resolved spectroscopy. A typical measurement scheme shines an extreme ultraviolet (EUV)
light pulse, in the form of an attosecond pulse train (APT) or a single attosecond pulse (SAP),
together with a synchronized infrared (IR) laser pulse, on an atomic or molecular target, and
measures the photoions, photoelectrons, or photoabsorption of the EUV. The adjustable time delay
between the EUV and
the IR gives information of the dynamics of the target. For the SAP in the x-ray regime,
this technique has been used to study the emission of Auger electrons~\cite{drescher} in the
so-called ``streaking'' model. Down to the EUV energy, a similar scheme was used to
time-resolve autoionization where the IR primarily just depleted the resonances excited by the
EUV~\cite{wang, gilbertson}. For a SAP at even lower energies, a series of bound states can
be excited at once. These states can then be ionized by the IR through different quantum
pathways to exhibit interference patterns~\cite{mauritsson}, or dressed by the IR temporarily
to exhibit the ac Stark shift~\cite{chini}.

The studies of coupled autoionizing states (AIS) by lasers have been carried out over the past
30 years theoretically~\cite{lambropoulos, bachau, madsen, themelis} and
experimentally~\cite{karapanagioti}. In these earlier investigations, long pulses were used.
For a typical AIS with a lifetime up to tenths of femtoseconds, the coupling lights were
considered to be monochromatic, and the measured spectrum was obtained by scanning the photon
energy over the widths of the resonances. Not until a few years ago was the same coupling scheme
extended to the EUV energy range with femtosecond pulses, where transient absorptions
were measured or calculated~\cite{loh, gaarde, tarana}. Very recently, with the emerging
attosecond pulses, the timescale of the coupling has been pushed shorter than the decay
lifetime of an AIS~\cite{wang, gilbertson} for direct observation in the time domain. In the
same scheme, numerous new spectroscopic features were suggested by simulations~\cite{chu11,
chu12, zhao, pfeiffer, chu12jpb, argenti}. Some of these features have
been shown recently by experiments with improved energy resolution~\cite{ott} of the
spectrometers. Further technological developments will likely bring more discoveries.

In this work, we focus on resonant laser coupling between two AISs, where an SAP passing
through the medium is strongly reshaped in its spectral and temporal distributions by
controlling the parameters of the laser and the medium. This reshaping of the SAP, which
behaves much beyond what can be described by the simple absorption rate or Beer's
law~\cite{ishimaru}, has not been investigated so far. In Ref.~\cite{gaarde}, the actual
propagation of EUV pulses was studied; however, those pulses with durations of tens of
femtoseconds are too narrow in bandwidth to demonstrate any meaningful modification in the
resonant spectrum. In this study, we look for how an SAP evolves in the medium, which cannot
be revealed by a single-atom absorption cross section.

For demonstration at the single-atom model, a 200 as EUV pulse and a time-delayed 9-fs laser
pulse are applied to the helium atom. The $2s2p(^1P)$ and $2s^2(^1S)$ AISs are coupled by a
540-nm laser with intensities between $10^{12}$ and $10^{13}$~W/cm$^2$, where the weak EUV
excites $2s2p$ from the ground state. For clarity, the pulses are not distinguished as the
pump or the probe since they just couple different sets of states. In the single-atom
calculation, the electronic wave function, photoelectron, and EUV absorption spectra are carried
out for a given set of pulses and atomic parameters. In order to extract the universal features
resulting from the coupling, we further derive simple analytic forms of the spectra for short
light pulses that are approximated by $\delta$ functions in time. This is a valid approximation
if the pulse durations are shorter than the atomic timescales, e.g., the 17 fs decay lifetime
of $2s2p$. The Fano $q$ parameter~\cite{fano} preserved in the final forms of the spectra
indicates how the population transfer between the AISs is controlled by the coupling and how
the wave packet evolves.

For the macroscopic model, the pulses are allowed to propagate through a 2 mm helium gas with
number density $8\times 10^{-3}$~cm$^{-3}$ (or a pressure of 25~Torr at room temperature). The
detected EUV spectra as the pulses exit the medium are calculated. We found that at certain
coupling and gas conditions, an enhancement of more than 50\% of the incident light at the
resonance is produced in the transmission spectrum. This enhancement peak in the frequency
domain persists through the propagation where the surrounding background decays along the
light path. This demonstrates a remarkable extension of the nonlinear optical control to the
attosecond time regime.

%%%%%%%%%%%%%%%%%%%%%%%%%%%%%%%%%%%%%%%%%%%%%%%%%%%%%%%%%%%%%%%%%%%%%%%%%%%%%%%%%%%%%%%%%%%%%%%
\section{Model}\label{sec:model}

%%%%%%%%%%%%%%%%%%%%%%%%%%%%%%%%%%%%%%%%%%%%%%%%%%%%%%%%%%%%%%%%%%%%%%%%%%%%%%%%%%%%%%%%%%%%%%%
\subsection{Single-atom wave function}\label{sec:model_wave}

\begin{figure}[tbp]
\centering
\includegraphics[width=0.50\textwidth]{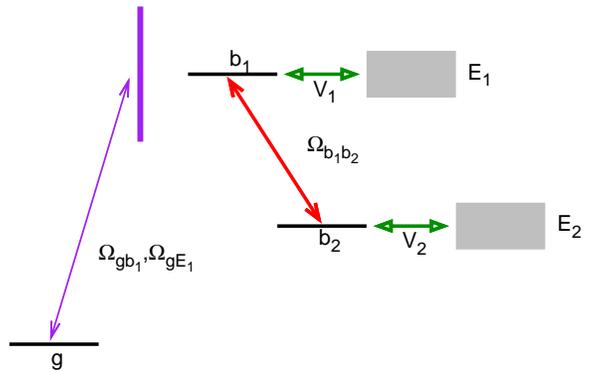}
\caption{(Color online) Coupling scheme used in the model. The thick red arrow is the laser
coupling. The thin purple arrow represents the EUV transition from the ground state to the
$|b_1\rangle$-$|E_1\rangle$ resonance, where the bandwidth of the EUV pulse, indicated by
the vertical purple line, widely covers the whole lineshape of the resonance. The green
arrows with hollow heads are the configuration
interactions responsible for the autoionization.}
\label{fig:scheme}
\end{figure}

In this section, we derive the time-dependent total wave function of a three-level autoionizing
system coupled by an EUV pulse and a laser pulse. Two AISs, composed by the bound parts
$|b_1\rangle$ and $|b_2\rangle$, and the associated background continua $|E_1\rangle$ and
$|E_2\rangle$, respectively, are coupled by the laser, while the $|b_1\rangle$-$|E_1\rangle$
resonance is coupled to the ground state $|g\rangle$ by the EUV, both by dipole transitions. The
coupling scheme can be either $\Xi$ type or $\Lambda$ type. A schematic of the system
is in Fig.~\ref{fig:scheme}. The pulses are linearly polarized in
the same direction and collinearly propagated. The EUV field is expressed by
$E_X(t) = F_X(t) e^{i\omega_Xt} + F_X^*(t) e^{-i\omega_Xt}$ where $F_X(t)$ is the envelope
and $\omega_X$ is the central frequency of the pulse. The pulse envelope is generally a
complex function, which includes all the phase factors in addition to the carrier-frequency
terms. The electric field $E_L(t)$ of the laser pulse, however, is kept in the exact form since
it may be a few-cycle pulse. The total wave function is in the general form of
\begin{align}
|\Psi(t)\rangle &= e^{-iE_gt} c_g(t) |g\rangle \nonumber\\
&+ e^{-i(E_g+\omega_X)t} \left[ c_{b_1}(t) |b_1\rangle + \int{c_{E_1}(t) |E_1\rangle dE_1} \right. \nonumber\\
& \left. + c_{b_2}(t) |b_2\rangle + \int{c_{E_2}(t) |E_2\rangle dE_2} \right].
\label{eq:Psi}
\end{align}

In solving the
time-dependent Schr\"{o}dinger equation (TDSE), we make the following approximations: (1) All
the second-order (two-electron) dipole matrix elements, except the resonant excitation
$\langle b_1 |\mu| g \rangle$ (e.g., the transition between $2s2p$ and $1s^2$), are neglected,
i.e., $\langle E_2|\mu|b_1 \rangle = \langle b_2|\mu|E_1 \rangle = 0$. (2) The rotating wave
approximation (RWA) is applied for the EUV coupling (but not the laser coupling) for its high
frequency and low intensity. (3) The matrix elements of the Hamiltonian involving $|E_1\rangle$
and $|E_2\rangle$ are independent of energy, where we are only concerned with the energy ranges
near the resonances. This means that the target structure is given in terms of the resonance
energies $E_{b_1}$ and $E_{b_2}$, the widths $\Gamma_1$ and $\Gamma_2$, and the $q$ parameter of the
$|b_1\rangle$-$|E_1\rangle$ resonance, which together constitute the basic Fano
lineshapes~\cite{fano}. Note that, with our first approximation, $q$ cannot be defined for
the $|b_2\rangle$-$|E_2\rangle$ resonance.

Within the prescribed truncated space, the TDSE for the total wave function in
Eq.~(\ref{eq:Psi}) is reduced to the coupled equations of the coefficients therein, where
these coefficients as functions in time should be calculated exactly with the given initial
conditions. However, the coefficients associated with the continua,  $E_1$ and $E_2$, make
the calculation less tractable. Thus, we adopt the adiabatic elimination of the continua by
assuming that they change much more slowly than the bound states. Then, the approximation
$\dot{c}_{E_1}(t)=\dot{c}_{E_2}(t)=0$ in the coupled equations gives
\begin{align}
c_{E_1}(t) &= \frac{1}{\delta_1-\Delta E_1} \left[ -F_X^*(t)D_{gE_1}^*c_g(t) + V_1 c_{b_1}(t) \right] \label{eq:adia1}\\
c_{E_2}(t) &= \frac{1}{\delta_2-\Delta E_2} V_2 c_{b_2}(t), \label{eq:adia2}
\end{align}
where $\delta_1 \equiv E_X-E_{b_1}$ and $\delta_2 \equiv E_X-E_{b_2}$ are the detuning of the
EUV relative to the transitions from the ground state to the AISs, $\Delta E_1 \equiv E_1 - E_a$
and $\Delta E_2 \equiv E_2 - E_b$ are the continuous energies relative to the resonance energies,
the $D$s are dipole matrix elements, and $V_1 \equiv \langle E_1 |H| b_1 \rangle$ and
$V_2 \equiv \langle E_2 |H| b_2 \rangle$ are the strengths of the configuration interaction.
In return, the coupled equations for the bound state coefficients are reduced to satisfy
\begin{align}
i\dot{c}_g(t) =& -F_X(t)D_{gb_1}^*\lambda c_{b_1}(t) -i\left|F_X(t)\right|^2j_{gg}c_g(t) \label{eq:cg}\\
i\dot{c}_{b_1}(t) =& -F_X^*(t)D_{gb_1}\lambda c_g(t) \notag\\
&- (\delta_1+i\kappa_1)c_{b_1}(t) - E_L(t)D_{b_1b_2}c_{b_2}(t) \label{eq:cb1}\\
i\dot{c}_{b_2}(t) =& -(\delta_2+i\kappa_2)c_{b_2}(t) - E_L(t)D_{b_1b_2}^* c_{b_1}(t), \label{eq:cb2}
\end{align}
where $\kappa_1 \equiv \Gamma_1/2$ and $\kappa_2 \equiv \Gamma_2/2$ are half widths of the
resonances, $\lambda
\equiv 1-i/q$ represents the transition from the ground state to $|b_1\rangle$ and
$|E_1\rangle$ jointly, and $j_{gg} \equiv \pi|D_{gE_1}|^2$ is proportional to the laser
broadening. The dipoles ($D$) in this work are all real numbers since real orbitals are used
for the bound states and standing waves are used for the continua. With
Eqs.~(\ref{eq:cg})-(\ref{eq:cb2}), the bound state part of the wave function is carried out. The
procedure so far is identical to what has been employed in the early studies concerning
coupled AISs~\cite{bachau, madsen, themelis}, except that we do not apply RWA on the laser
pulse.

For pulses that are long compared to the resonance lifetime, the pulse bandwidth is narrow,
thus each measurement gives the ionization yield at just one single photon energy in the
spectrum. In this case, calculation of the ground-state population $c_g(t)$ described above
would be enough for carrying out the ionization spectra. However, in the present study, the
bandwidth of the attosecond pulse is broad and can excite the whole AIS, including a
significant fraction of the background continuum. The continuum part of the wave function is
needed in determining the electron or absorption spectra in a single measurement. Here, the
coefficients of the continuum part are retrieved from the original coupled equations
\begin{align}
i\dot{c}_{E_1}(t) =& (E_1-E_X)c_{E_1}(t) + V_1 c_{b_1}(t) \notag\\
& - F_X^*(t)D_{gE_1}^*c_g(t) \label{eq:cE1}\\
i\dot{c}_{E_2}(t) =& (E_2-E_X)c_{E_2}(t) + V_2 c_{b_2}(t). \label{eq:cE2}
\end{align}
The retrieval of the continua can be viewed as a ``correction'' after an iteration from the
preliminary forms in Eqs.~(\ref{eq:adia1}) and (\ref{eq:adia2}). Now, the time-dependent
total wave function in the form of Eq.~(\ref{eq:Psi}) is completely solved.

%%%%%%%%%%%%%%%%%%%%%%%%%%%%%%%%%%%%%%%%%%%%%%%%%%%%%%%%%%%%%%%%%%%%%%%%%%%%%%%%%%%%%%%%%%%%%%%
\subsection{Single-atom spectrum}\label{sec:model_spect}

In the following, the electron energy spectrum and photoabsorption spectrum for a single atom are
defined by the probability density per unit energy for emitting an electron and for absorbing
a photon, respectively. In macroscopic cases where propagation is taken
into account, the spectroscopy is expressed in terms of the transmitted light intensity
versus the photon energy.

%%%%%%%%%%%%%%%%%%%%%%%%%%%%%%%%%%%%%%%%%%%%%%%%%%%%%%%%%%%%%%%%%%%%%%%%%%%%%%%%%%%%%%%%%%%%%%%
\subsubsection{Photoelectrons}\label{sec:model_elec}

In standard scattering theory, the atomic scattering waves for momentum $\vec{k}$ are given by
\begin{equation}
\psi_{\vec{k}}(\vec{r}) = \sqrt{\frac{2}{\pi k}} \frac{1}{r} \sum_{lm} i^l e^{i \eta_l} u_l(kr)
Y_{lm}(\hat{r}) Y_{lm}^*(\hat{k}) \label{eq:mom}
\end{equation}
in the energy-normalized form, where $u_l$ are real standing wave radial functions, and
$\eta_l$ are the scattering phase shifts. Both $u_l$ and $\eta_l$ are determined by the
atomic potential. If one detects photoelectrons, the momentum distribution will be the
projection of the total wave function at a large time onto the scattering waves, i.e.,
\begin{equation}
P(\vec{k}) = \lim_{t\to\infty} \left| \left\langle\psi_{\vec{k}} | \Psi(t)\right\rangle \right|^2.
\end{equation}
Likewise, if one measures the energy of the photoelectrons, the spectrum is given by $P(E) =
\lim_{t\to\infty} \left|\left\langle \psi_E | \Psi(t)\right\rangle \right|^2$, where
$|\psi_E\rangle$
are the scattering waves for energy $E$. Remember that $|E_1\rangle$ and $|E_2\rangle$
are conventionally chosen as standing waves, which contains both the incoming and the outgoing
components. At large times, since physically the AISs will decay to the continuum, the wave
packet will not have incoming components, i.e., the projection onto
scattering waves is the same as the projection onto standing waves. Thus, the photoelectron
spectra corresponding to the two resonances are simply
\begin{align}
P_1(E_1) = \lim_{t\to\infty} \left| c_{E_1}(t) \right|^2, \notag\\
P_2(E_2) = \lim_{t\to\infty} \left| c_{E_2}(t) \right|^2, 
\end{align}
which require us to calculate the total wave function for a physical time much longer than the
decay lifetimes, until $c_{E_1}(t)$ and $c_{E_2}(t)$ stop evolving other than an oscillating
phase.

%%%%%%%%%%%%%%%%%%%%%%%%%%%%%%%%%%%%%%%%%%%%%%%%%%%%%%%%%%%%%%%%%%%%%%%%%%%%%%%%%%%%%%%%%%%%%%%
\subsubsection{Photoabsorption}\label{sec:model_abs}

In a given external field, the total energy absorbed by the atom is
\begin{equation}
\Delta U = \int_0^{\infty} { \omega S(\omega) d\omega },
\end{equation}
where the response function $S(\omega)$ for $\omega > 0$ represents the absorption
probability density, and is proportional to the pulse intensity in the frequency
domain~\cite{gaarde}. Thus, $P(E)$ and $S(\omega)$ have the same dimension and can be
directly compared. The absorption cross section $\sigma(\omega)$ is related to
$S(\omega)$ by
\begin{equation}
\sigma(\omega) = \frac {4\pi\alpha\omega S(\omega)} {\left| \tilde{E}(\omega)
\right|^2}, \label{eq:xsec}
\end{equation}
where $\alpha$ is the fine structure constant, and $\tilde{E}(\omega)$ is the Fourier
transform of the electric field. By considering dipole interaction between
the electric field and the atom, $S(\omega)$ is given by
\begin{equation}
S(\omega) = -2 \text{Im} \left[ \tilde{\mu}(\omega) \tilde{E}^*(\omega) \right],
\label{eq:S}
\end{equation}
where $\tilde{\mu}(\omega)$ is the Fourier transform of the dipole moment.

With the wave function in Eq.~(\ref{eq:Psi}), the dipole moment is given by
\begin{equation}
\mu(t) = e^{i\omega_Xt} u_X(t) + u_L(t) + \text{c.c.}, \label{eq:mu}
\end{equation}
where the carrier frequency of the EUV has been factored out, and
\begin{align}
u_X(t) &= D_{gb_1}\lambda^*c_{b_1}^*(t)c_g(t) - iF_X(t)j_{gg}\left|c_g(t)\right|^2, \label{eq:uX}\\
u_L(t) &= D_{b_1b_2} c_{b_2}^*(t) c_{b_1}(t). \label{eq:uL}
\end{align}
Note that this model aims to deal with both short and long pulse lasers. In the case of a
few-cycle laser, the Fourier transform $\tilde{\mu}(\omega)$ at very low frequencies may
contain the contributions from both $u_L(t)$ and its complex conjugate, where the separation
of the carrier frequency and the envelope is not beneficial; we thus do not employ this
separation for the laser pulse in our study.

In a typical ultrafast EUV-plus-IR experiment, the electron or absorption spectra measured
over variable time delays between the two pulses are the most common. Photoions are also
usually measured, but they do not carry additional information. To study dynamic behavior
of autoionization directly in the time domain, light pulses shorter than the typical decay
lifetime are necessary. By now few-cycle
lasers of durations less than 10 fs are quite common. In the case where both pulses
are significantly shorter than any other atomic timescale, major features of the dynamics
can be recovered by assuming the pulses are $\delta$-functions in time. With this
approximation, we are able to reduce the electron and absorption spectra to very simple
forms in terms of the pulse areas of the EUV and the laser, the time delay, and the
$q$ parameter, where the detailed derivations are shown in the appendix.
The modification brought by the laser will have maximum effects when the laser follows
the EUV immediately and its pulse area is $(4N+2)\pi$ for integer $N$. In this case the
photoelectron distribution flips about the resonance energy. For the light, the
absorption line changes to the emission line. These dramatic changes gradually
die down when the time delay increases in terms of the decay lifetime of the observed
resonance. This $\delta$-pulse analysis will serve to qualitatively explain the result of
realistic numerical calculation later in this article.

%%%%%%%%%%%%%%%%%%%%%%%%%%%%%%%%%%%%%%%%%%%%%%%%%%%%%%%%%%%%%%%%%%%%%%%%%%%%%%%%%%%%%%%%%%%%%%%
\subsection{Propagation of light}\label{sec:model_prop}

We have shown how to model the resonance shape of an AIS changed by the strong coupling laser
in an atomic system. In principle, both the electron and absorption spectra carry the same
essential information, depending on the resonance structure of the states and the laser
parameters. However, in electron measurements, high precision spectroscopy is usually harder
to achieve since the  lineshape of most Fano resonances requires meV energy resolution. In
contrast, such resolution is easier with the transient absorption spectroscopy, as
exemplified in a recent report~\cite{ott}. Thus, in the present work we will put emphasis
on absorption spectroscopy.

In a linear medium, the absorption of light is commonly described by Beer's law as
\begin{equation}
T(\omega) = T_0(\omega) \text{exp} \left[ -\rho\sigma(\omega)L \right], \label{eq:beer}
\end{equation}
where $T_0(\omega)$ and $T(\omega)$ are light intensities at the entrance and at the exit of
the medium, respectively, $\rho$ is the number density of the particles, $\sigma(\omega)$ is
the absorption cross section, and $L$ is the travel distance of light in the medium. This
form requires that the cross section has no spatial and temporal dependence and is only a
function of energy, and the transmission of light simply decays exponentially with its
traveling distance and the gas density. However, in the presence of intense ultrashort IR
pulses, the response of the medium can be more complicated, which should be
carried out by propagating the exact electric field in the medium.

In the present work, by assuming that the laser is loosely focused, the electric field is
a function of time $t$ and the spatial coordinate in the propagation direction $z$ only,
where the dependence on the transverse direction
is neglected. We assume that the propagation is only forward in $z$ and approximately at the
speed of light in vacuum, $c$.  By expressing the time in the moving frame $t' \equiv t-z/c$,
the Maxwell equation is reduced to
\begin{equation}
\frac{\partial E(z,t')}{\partial z} = -\frac{\rho}{c\epsilon_0} \frac{\partial
\mu(z,t')}{\partial t'}, \label{eq:maxwell}
\end{equation}
where $\rho$ is the gas density. Because the EUV and laser frequencies are widely separate,
they can be propagated separately. For the EUV, the carrier oscillation
terms in field and in dipole are factored out, and the propagation for the field envelope is
\begin{equation}
\frac{\partial F_X(z,t')}{\partial z} = -\frac{\rho}{c\epsilon_0} \left[ \frac{\partial
u_X(z,t')}{\partial t'} + i\omega_Xu_X(z,t') \right]. \label{eq:maxwell_env}
\end{equation}
For the laser, the propagation simply follows Eq.~(\ref{eq:maxwell}) where $\mu(z,t')$ is
replaced by $u_L(t')+\text{c.c.}$. The dipole oscillations corresponding to the EUV and laser fields
are determined by Eqs.~(\ref{eq:uX}) and (\ref{eq:uL}), respectively, and these dipoles
determine how the pulses evolve along $z$ by Eqs.~(\ref{eq:maxwell_env}) and (\ref{eq:maxwell}).
The TDSE for the single-atom wave function and the Maxwell equation for light propagation are
executed sequentially until the pulses exit the medium.  

%%%%%%%%%%%%%%%%%%%%%%%%%%%%%%%%%%%%%%%%%%%%%%%%%%%%%%%%%%%%%%%%%%%%%%%%%%%%%%%%%%%%%%%%%%%%%%%
\section{Results and Analysis}\label{sec:results}

In the following, energy and time are expressed in electron volts (eV) and femtoseconds (fs),
respectively, and the signal intensities in spectra are expressed in atomic units (a.u.),
unless otherwise specified.

%%%%%%%%%%%%%%%%%%%%%%%%%%%%%%%%%%%%%%%%%%%%%%%%%%%%%%%%%%%%%%%%%%%%%%%%%%%%%%%%%%%%%%%%%%%%%%%
\subsection{Results for a single atom}\label{sec:results_atom}

In this study, the characteristics of two strongly coupled AISs are demonstrated for neutral
helium, the simplest atomic system possessing electron correlation effects. The lowest few
AISs in helium are separated from one another on the scale of eVs and are easily coupled
by lasers in the near IR to the visible-light energy range. Here we consider a laser pulse
[wavelength $\lambda_L=540$~nm, full width at half maximum (FWHM) duration $\tau_L=9$~fs,
peak intensity $I_L=2I_0$ where $I_0\equiv1$~TW/cm$^2$ hereafter for convenience]
to resonantly couple the
$2s2p(^1P)$ and $2s^2(^1S)$ resonances, and a weak EUV SAP (central photon energy
$\omega_X=60$~eV, FWHM duration $\tau_X=200$~as, peak intensity 10$^{10}$ W/cm$^2$) to
excite $2s2p$ from the ground state by one-photon transition. Both pulse envelopes are in the
sine-squared form. The $\Lambda$ coupling
scheme was chosen so that the binding energies of the two AISs are moderately high to prevent
ionization by the laser. The time delay $t_0$ is adjustable, which is defined by the time
between the two pulse peaks, and it is positive when the EUV comes first. The broadband EUV
covers roughly from 50 to 70 eV and can in principle excite
many resonances at once, but due to the choice of the laser wavelength and its relatively
narrow bandwidth of about 200 meV, the coupling between the two AISs specified above remains
our focus, where it will only be slightly
disturbed by the presence of other states if they are not totally negligible.

\begin{figure}[tbp]
\centering
\includegraphics[width=0.50\textwidth]{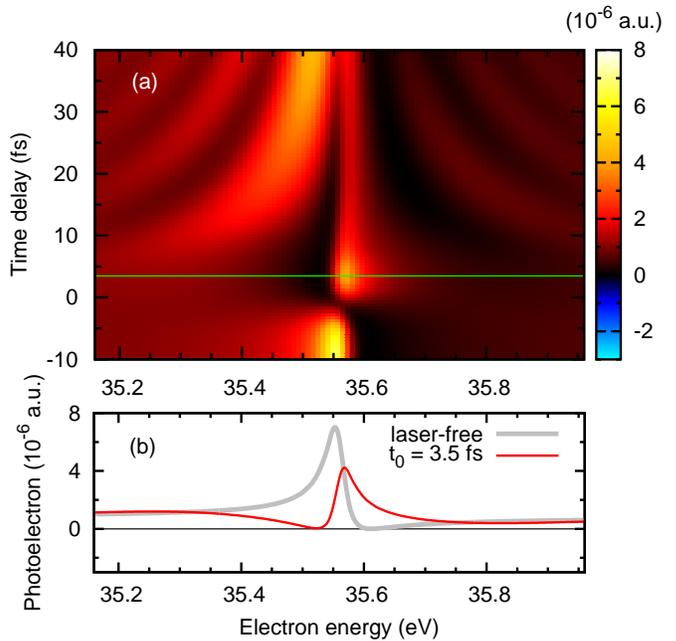}
\caption{(Color online) Time-delayed photoelectron spectra near the $2s2p$ resonance excited
by a 200 as EUV with a 9 fs laser coupling, calculated by the single-atom model.
(a) The spectra between $t_0=-10$ and 40~fs. The spectra for positive delays are to be
compared with those in the $\delta$-pulse analysis in Fig.~\ref{fig:shortP}. (b) The
spectrum for $t_0=3.5$~fs, which corresponds to the green line in panel (a). At this time
delay, the resonance profile flips horizontally from the original Fano lineshape,
which is shown by the gray curve. }
\label{fig:singleP}
\end{figure}

\begin{figure}[tbp]
\centering
\includegraphics[width=0.50\textwidth]{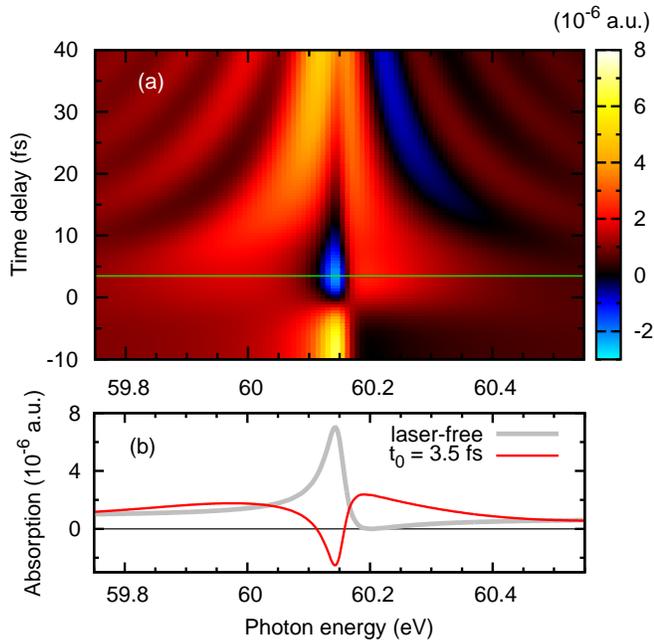}
\caption{(Color online) Same as Fig.~\ref{fig:singleP} but for the photoabsorption spectra.
The spectra for positive delays are to be compared with the spectra in the $\delta$-pulse
analysis in Fig.~\ref{fig:shortS}. The resonance profile for $t_0=3.5$~fs flips vertically
from the original Fano lineshape shown by the gray curve. Note that the negative signals at
$t_0=3.5$~fs around the resonance energy represent the emission of light. In contrast, the
photoelectron spectra in Fig.~\ref{fig:singleP} do not have negative signals.}
\label{fig:singleS}
\end{figure}

The time-delayed photoelectron and photoabsorption spectra near the $2s2p$ resonance energy,
calculated by the single-atom model, are shown in Figs.~\ref{fig:singleP}(a) and
\ref{fig:singleS}(a), respectively, for the time delay ranging from $-10$~fs to 40~fs. Most
informative spectral features are included in this range. The resonant coupling laser with
intensity $2I_0$ has a pulse area of $1.3\times2\pi$, which means that the laser pulse,
in its whole, is able to incur Rabi oscillation for 1.3 cycles.

The spectra in Figs.~\ref{fig:singleP}(a) and \ref{fig:singleS}(a), calculated with
realistic pulse shapes, retain the main features shown in Figs.~\ref{fig:shortP} and
\ref{fig:shortS}. The latter two figures are derived analytically by assuming that the two
pulses are $\delta$-functions in time. The most obvious differences appear when the two pulses
overlap, where only the part of the laser after the EUV pulse is responsible for the Rabi
oscillation. Unlike the $\delta$-function short pulses, the realistic pulses bring their own
timescales into account. Thus, we define an ``effective pulse area'' by counting only the laser
duration after the EUV peak. For $t_0=3.5$~fs, the effective pulse area is $2\pi$, i.e., the
laser at that time delay supports perfectly a full cycle of Rabi oscillation.
In Figs.~\ref{fig:singleP}(b) and \ref{fig:singleS}(b), the spectra for $t_0=3.5$~fs are
plotted. The flipped Fano lineshapes discussed in Sec.~\ref{sec:model_abs} and in the
appendix are seen. In particular, the electron spectrum flips horizontally where the sign
of $q$ is changed, and the absorption spectrum has a upside-down image, where the
absorption peak in the laser-free spectrum points downward, and the signals below the zero
line are for the light emission. These ``flipped Fano lineshape'' features have been shown
in theoretical studies~\cite{chu11, chu12, zhao, chu12jpb, argenti} and observed in a recent
experiment~\cite{ott}. While some comprehensive simulations were done previously, the
analytical derivation in this work would provide a simple yet effective explanation that could
readily be applied.

Note that the AISs decay continuously once they are populated, no matter whether the laser
is present or not. For $t_0=3.5$~fs, at the time of the laser peak, the $2s2p$ state has
already partially autoionized, and the Rabi oscillation is able to affect only electrons
that are not autoionized yet. Furthermore, the decay of $2s^2$ also ``leaks'' some electrons
resulting from Rabi oscillation, by emitting photoelectrons with its own decay
lifetime of 5.3~fs. These factors reduce the distortions that the laser would have imposed on
the original Fano lineshape, in both the electron and absorption spectra. As the time delay
increases, more electrons are autoionized before the laser, and less are coupled by the laser
to participate in the Rabi oscillation, and the resonance profile gradually approaches the
original Fano lineshape.

\begin{figure}[tbp]
\centering
\includegraphics[width=0.50\textwidth]{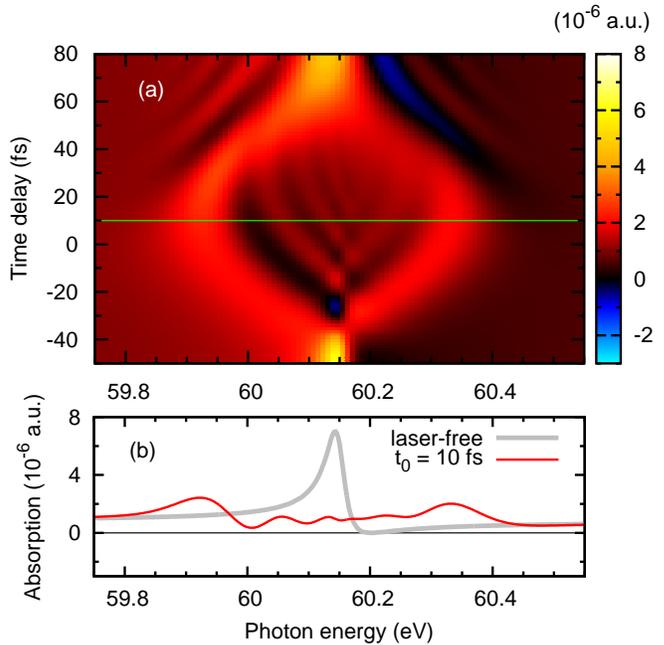}
\caption{(Color online) Time-delayed photoabsorption spectra of a 200 as EUV with a 50 fs
dressing laser in the single-atom model. (a) The spectra between $t_0=-50$~fs and 80~fs.
The EIT condition is controlled by the time delay between the two pulses. (b) The spectrum for
$t_0=10$~fs shows the largest separation in the Autler-Townes doublet, which matches the
Rabi frequency at the laser peak.}
\label{fig:singleA}
\end{figure}

While similar three-level coupling schemes have been used for the electromagnetically induced
transparency (EIT) effect~\cite{harris, fleischhauer}, the spectral features in a typical
EIT setup, such as the transparency line or Autler-Townes doublet~\cite{autler}, are not
recovered by our photoabsorption calculation in Fig.~\ref{fig:singleS}. Evidently, the
approximation with $\delta$-function pulses breaks down in the long dressing field. To
differentiate our mechanism from EIT, we make an additional calculation by keeping all the
parameters, but the laser duration is extended from 9~fs to 50~fs. The resultant spectra are
plotted in Fig.~\ref{fig:singleA}. It shows that the long laser pulse splits the resonance
peak into two peaks, with the maximum separation 0.45~eV at $t_0=10$~fs. The separation is
exactly the Rabi frequency at the laser peak, where the Autler-Townes doublet is
reproduced. As the time delay changes from this optimal value, the separation decreases, and
finally drops to 0 for $t_0<-40$~fs or $t_0>70$~fs. It shows that contrary to short dressing
pulses, long dressing pulses with many Rabi cycles creates the condition satisfying the
``dressed state'' picture behind the EIT phenomenon, and the split states appear as expected.
By controlling the overlap between the pulses, the doublet can be ``turned on'' or ``turned
off.'' This is the theme of some recent studies applying dressing lasers of tens to hundreds
of femtoseconds in an EIT scheme while measuring the time-delayed EUV
transmission~\cite{loh, tarana, glover}.

%%%%%%%%%%%%%%%%%%%%%%%%%%%%%%%%%%%%%%%%%%%%%%%%%%%%%%%%%%%%%%%%%%%%%%%%%%%%%%%%%%%%%%%%%%%%%%%
\subsection{Results for a gaseous medium}\label{sec:results_gas}

In Sec.~\ref{sec:results_atom}, the photoabsorption spectrum reveals the appearance of emission
line at the resonance for $t_0=3.5$~fs. However, it is physically impossible for the light at
certain frequency to gain intensity indefinitely when
passing through a medium. A realistic pulse has no singularity in the frequency domain. Thus, by
intuition, the single-atom result does not represent what will actually be measured in a
dense gaseous medium. With the same EUV and laser pulses used in Sec.~\ref{sec:results_atom}, we
consider a gaseous medium made of non-interacting helium atoms with number density
$\rho=8\times10^{17}$~cm$^{-3}$ (equivalent to pressure of 25~Torr at room temperature) and
thickness $L=2$~mm, and calculate the transmitted light spectra at the exit of the medium.

\begin{figure}[tbp]
\centering
\includegraphics[width=0.50\textwidth]{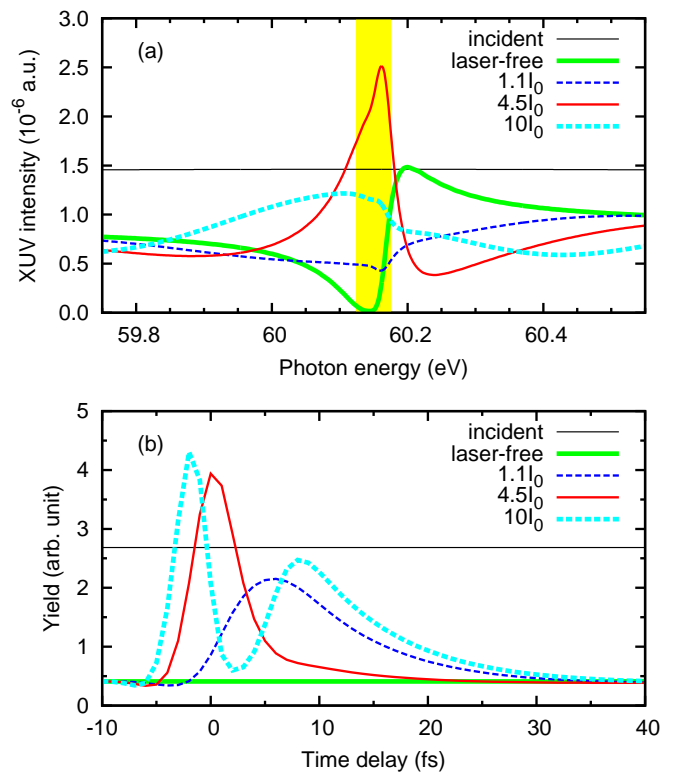}
\caption{(Color online) Transmission of EUV through a helium gaseous medium dressed by a 9 fs
laser of peak intensities ($I_L$) of 0
(laser free), $1.1I_0$, $4.5I_0$, and $10I_0$. The incident light is plotted as a reference.
(a) Transmitted EUV profiles for overlapping pulses ($t_0=0$), where the effective pulse areas
of the laser are 0, $\pi$, $2\pi$, and $3\pi$. The flipping of the resonance profile is
predicted as the main feature by the single-atom case, although more structures are seen here.
The signal at the resonance for $I_L=4.5I_0$ is enhanced significantly, with more than 50\%
increase from the incident light. This measurable quantity can only be obtained by applying
field propagation. (b) Total photon signals (yield) gathered in the 50-meV energy range at the
resonance [indicated in panel (a) by the yellow vertical stripe] as functions of the time delay.
This energy range complies with the spectrometer resolution under current technology. The yield
for each laser intensity oscillates with the time delay, which shows a high degree of control
in the time domain.}
\label{fig:propXY}
\end{figure}

The transmitted EUV is plotted in Fig.~\ref{fig:propXY} for three laser intensities,
$I_L=1.1I_0$, $4.5I_0$, and $10I_0$, where $I_L$ is the peak intensity. They safely fall into
the intensity range considered by our model.
The EUV spectra with overlapping pulses fixed at $t_0=0$ are plotted in Fig.~\ref{fig:propXY}(a).
The effective pulse areas for the three intensities are $\pi$, $2\pi$, and $3\pi$, which are
responsible for the half, one, and one-and-half cycles of Rabi oscillation. For $1.1I_0$, the
spectrum is mostly flat, indicating that while the laser moves the electron population from
$2s2p$ to $2s^2$, the $2s2p$ AIS ``disappears,'' leaving only the directly ionized
photoelectrons in the vicinity without the hint of autoionization. For $4.5I_0$, the laser
drives the electrons back to $2s2p$ with a phase shift of $\pi$. The emission line that has been
discussed in the single-atom picture is observed here, where the spectral peak appears to be
more than 50\% of the signal intensity of the incident pulse, which is effectively a
partial enhancement of the SAP passing through the gas. When the laser intensifies
further to $10I_0$, the Rabi oscillation drives the electrons away from $2s2p$ again; the
enhancement is gone, and the spectrum looks flat other than a slight bump. These three laser
intensities demonstrate how the number of cycles in the Rabi oscillation controls the phase
of the $2s2p$ AIS, which then determines the resonance lineshape at the end of the
autoionization.

The presentation in Fig.~\ref{fig:propXY}(a) is for the overlapping pulses, i.e., the Rabi
oscillation happens within the laser duration at the beginning of the 17 fs decay of $2s2p$.
It is a simplified case where  autoionization occurs approximately after the population
transfer by the laser is done. However, for larger time delays, the laser comes after
$2s2p$ decays for some time and only affects the later part of the
decay. The total wave packet is the coherent sum of autoionization before the
laser, and the quantum path going through the Rabi oscillation. In Fig.~\ref{fig:propXY}(b),
the total signal yields in the 50 meV range around the $2s2p$ resonance for different
intensities and time delays are plotted. This plot is equivalent to a measurement of the EUV
signal at the resonance energy with a spectrometer of 50 meV resolution in order to
investigate the enhancement. For each intensity, the yield oscillates with the
time delay. As seen in the figure, $4.5I_0$ is no longer the definite optimal
intensity to enhance the transmitted SAP resonance. For example, when the $10I_0$ pulse is
at $t_0=-2$~fs, the enhancement is higher than what can be achieved by the $4.5I_0$ pulse.
This is because even when both intensities have the same effective pulse area, the $10I_0$
pulse accomplishes the Rabi oscillation in a slightly earlier stage in the decay of $2s2p$,
where less electrons escape its influence by autoionization. It is very remarkable that this
partial enhancement of a SAP is controlled within an energy scale of about 100~meV in the
form of resonance lineshape, and within a timescale of about 1~fs in the form of time delay.

\begin{figure}[tbp]
\centering
\includegraphics[width=0.50\textwidth]{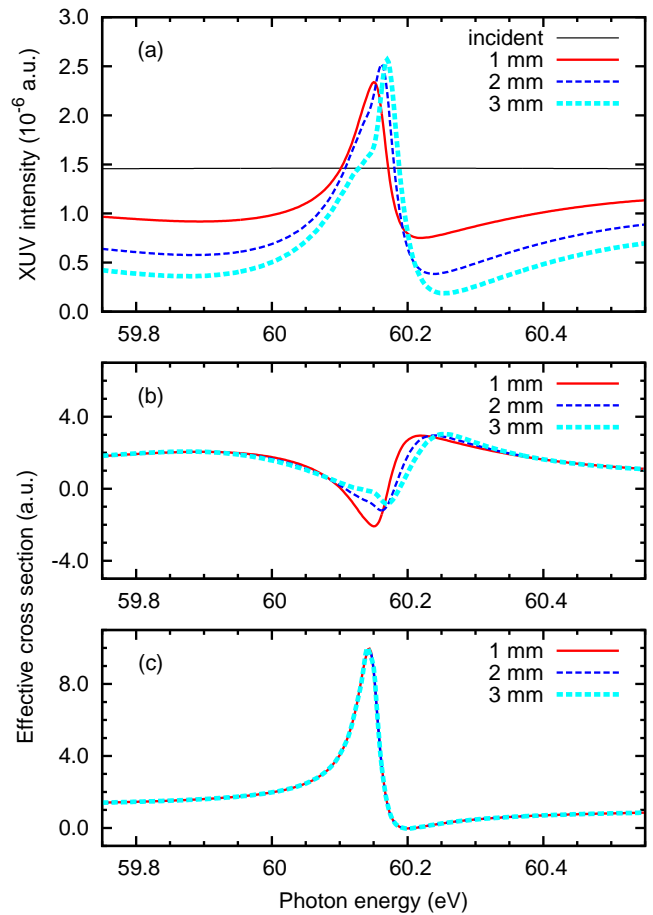}
\caption{(Color online) (a) EUV transmission profiles for propagation lengths $L=1$, 2,
and 3~mm in the gaseous medium with the overlapping 9 fs, $4.5I_0$ dressing laser pulse.
When the dressing laser is turned on, the enhanced peak persists while the surrounding
background attenuates. Panels (b) and (c) show effective absorption cross sections
$\sigma_{\text{eff}}(\omega)$ for the same propagation lengths as in panel (a) with and
without the dressing laser, respectively. With the dressing
laser, $\sigma_{\text{eff}}(\omega)$ changes with $L$ at the resonance but stays the same in the
outside region. Without the dressing field, $\sigma_{\text{eff}}(\omega)$ over the whole energy
range does not change with $L$, where Beer's law applies.
The contrast between the cases with and without the dressing laser demonstrates a strong
nonlinear response of the gas to the laser field which is reflected in the transmission of
the attosecond pulse.}
\label{fig:propL}
\end{figure}

It has been demonstrated in Fig.~\ref{fig:propXY}(a) that while the resonance lineshape of
the EUV pulse is controlled by the laser coupling, the spectral profile away from the
resonance energy always drops in light transmission spectroscopy. It is imaginable
that the overall light attenuates when propagating in the gas.
It is thus very intriguing to know how the exceptional enhanced resonance peak holds its
shape and strength in the propagation. In Fig.~\ref{fig:propL}, we show the transmitted EUV
spectra at different distances along the beam path with the 9-fs, $4.5I_0$ dressing pulse
at 0 time delay. The transmitted EUV intensities and the corresponding effective absorption
cross sections, defined by [see Eq.~(\ref{eq:beer})]
\begin{equation}
\sigma_{\text{eff}}(\omega) \equiv -\frac{1}{\rho L} \ln{\frac{T(\omega)}{T_0(\omega)}}, \label{eq:xsec_eff}
\end{equation}
are shown in Figs.~\ref{fig:propL}(a) and (b), respectively. With such laser parameters, the
``wings'' on the two sides of the resonance profile descend, but the enhancement peak is up
to about 50\% of the incident intensity, and roughly maintains its height from 1 to
3~mm. The persistence of the enhancement peak, apart from the exponential decay of the rest
of the signals along the propagation, displays an impressive nonlinear response of the gas
to the laser pulse. This violation to Beer's law is in a very selective energy range and is
sensitive to the parameters of the dressing laser. To elucidate the nonlinearity of the
dressing laser, $\sigma_{\text{eff}}(\omega)$
without the dressing field is plotted in Fig.~\ref{fig:propL}(c). In this case, the propagation
of the EUV actually follows Beer's law stated in Eq.~(\ref{eq:beer}), where the cross section
determined by the original Fano lineshape remains the same over the distance. The pulse over
the whole energy range is absorbed indifferently.

%%%%%%%%%%%%%%%%%%%%%%%%%%%%%%%%%%%%%%%%%%%%%%%%%%%%%%%%%%%%%%%%%%%%%%%%%%%%%%%%%%%%%%%%%%%%%%%
\section{Conclusion}\label{sec:conclusion}

A three-level autoionizing system coupled by a SAP and a time-delayed intense femtosecond
laser have been modeled. The photoelectron and photoabsorption spectra are responsive most
obviously to the laser intensity and its time delay. We analyze the spectra by assuming both
pulses are infinitely short, where the responses can be explained by the precise control of the
Rabi oscillation between the two AISs. The analysis also suggests that the phase change in the
wave packet during the Rabi oscillation is mapped out in the flipping of the Fano lineshape.
The model calculation with realistic pulse parameters also produces these significant
features. In contrast, with long dressing fields, these spectral features are replaced by
the Autler-Townes doublet, which is well studied in systems coupled by stable lasers. This
contrast demonstrates that the ultrashort coupling demolishes the dressed-state picture, and
the dynamics of the system is better viewed directly in the time domain.

Focusing on the response of the SAP light in this coupled system, we further incorporate the
single-atom response with the pulse propagation in the gaseous medium. The main controllable
features of the resonance profile shown by a single atom are well preserved in the gaseous medium.
By applying a laser pulse with a $2\pi$ effective pulse area to couple the $2s2p$ and $2s^2$
resonances in the dense helium gas, an enhanced peak at the $2s2p$ resonance shows up in the
transmitted SAP spectrum, where the signal intensity retains more than 50\% gain from the
entrance to the exit of the medium. At the same time, the signals away from the resonance
energy drops along the propagation as described by Beer's law. It can be viewed as the
pulse shaping of a SAP that is measurable in the energy domain. This result illustrates a
strong nonlinear manipulation of an SAP by an ultrafast dressing field, which would open new
possibilities of control of attosecond dynamics.

%%%%%%%%%%%%%%%%%%%%%%%%%%%%%%%%%%%%%%%%%%%%%%%%%%%%%%%%%%%%%%%%%%%%%%%%%%%%%%%%%%%%%%%%%%%%%%%
\begin{acknowledgments}
This work is supported in part by Chemical Sciences, Geosciences and Biosciences Division,
Office of Basic Energy Sciences, Office of Science, U.S. Department of Energy.
C.~D.~L. would also like to acknowledge the partial support of National Taiwan University
(Grant No. 101R104021 and No. 101R8700-2).
\end{acknowledgments}

%%%%%%%%%%%%%%%%%%%%%%%%%%%%%%%%%%%%%%%%%%%%%%%%%%%%%%%%%%%%%%%%%%%%%%%%%%%%%%%%%%%%%%%%%%%%%%%
\appendix
\section{Atomic response in the short-pulse limit}

In the short-pulse limit, we assume that the field envelopes are rectangular with given
carrier frequencies. As the durations approach zero, the envelopes are
\begin{align}
F_X(t) &= \frac{A_X}{2} D_{gb_1} \delta(t) \notag\\
F_L(t) &= \frac{A_L}{2} D_{b_1b_2} \delta(t-t_0), \label{eq:shortF}
\end{align}
where $A_X$ and $A_L$ are equivalent to the pulse areas for the EUV and laser, respectively,
and $t_0$ is the time delay between the pulse peaks, which is positive when the EUV comes
first. Since the intensity of the laser is not enough to excite the atom from the ground
state, we only consider the time delay $t_0 \geq 0$. The infinitely short pulses cut the time
into regions where the TDSE in each region can be solved analytically. The EUV is weak so that
perturbation is applied on its transition. In order to observe
the universal features of the electron and absorption spectra regardless of the atomic
parameters, the energy is scaled by $\epsilon \equiv 2(E-E_{b_1})/\Gamma_1$, where $\Gamma_1$
is the resonance width (see Sec.~\ref{sec:model}). Note that now $E_{b_1}$ represents either
the electron energy or the photon energy depending on whether we look at the electron or
absorption spectra. This scaling is the same as Fano's characterization of
resonances~\cite{fano}. The time delay is scaled by $\tau \equiv \Gamma_1 t_0/2$.

With some algebra, the electron spectrum is found to be given by
\begin{equation}
P(\epsilon) = \frac{\left|A_X\right|^2}{1+\epsilon^2} \left[ (q+\epsilon)^2 - ce^{-\tau} f_P(\epsilon)
\right], \label{eq:shortP}
\end{equation}
where
\begin{equation}
c \equiv 1-\cos(A_L/2) \label{eq:c_AL}
\end{equation}
and
\begin{equation}
f_P(\epsilon) \equiv 2(q+\epsilon) \left[ q\cos(\epsilon \tau) + \sin(\epsilon \tau) \right]
+ ce^{-\tau} (q^2+1). \label{eq:shortfP} \\
\end{equation}
Similarly, the absorption spectrum is given by
\begin{equation}
S(\epsilon) = \frac{\left|A_X\right|^2}{1+\epsilon^2} \left[ (q+\epsilon)^2 - ce^{-\tau} f_S(\epsilon) \right],
\label{eq:shortS}
\end{equation}
where
\begin{equation}
f_S(\epsilon) \equiv (q^2-1+2q\epsilon) \cos(\epsilon \tau) + \left[ 2q + (1-q^2) \epsilon \right]
\sin(\epsilon \tau). \label{eq:shortfS}
\end{equation}
Both spectra are composed of two terms. The first term is the original Fano lineshape defined
by the $|b_1\rangle$-$|E_1\rangle$ resonance alone, without any influence from the laser.
The second term decays exponentially with the time delay $\tau$, and is proportional to the
coefficient $c$ determined by the laser pulse area $A_L$ as defined by Eq.~(\ref{eq:c_AL}).
The minimum value of $c$ is 0, which corresponds to $A_L=0,4\pi,8\pi,...$, where the Rabi
oscillation has even number of cycles. In this condition, the electrons coming back to
$|b_1\rangle$ carry no additional phase, and the laser effectively does nothing. On the
contrary, the maximum value of $c$, 2, corresponds to $A_L=2\pi,6\pi,10\pi,...$ and odd number
of cycles. The returning electrons have additional phase of $\pi$. \newpage

\begin{figure}[htbp]
\centering
\includegraphics[width=0.50\textwidth]{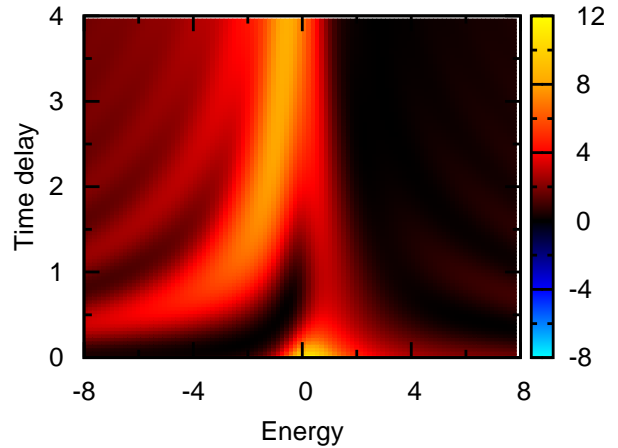}
\caption{Photoelectron spectra with infinitely short pulses, as described by
Eq.~(\ref{eq:shortP}). The energy is scaled by $\Gamma_1/2$, and the time delay is scaled by
$2/\Gamma_1$. The pulse area of laser is $A_L=2\pi$, which indicates a full cycle of Rabi
oscillation by the laser coupling. When the time delay is large, the spectrum approaches the
original Fano lineshape. When the time delay is 0, the spectrum is close to a
horizontally flipped image of the original Fano lineshape.}
\label{fig:shortP}
\end{figure}

\begin{figure}[htbp]
\centering
\includegraphics[width=0.50\textwidth]{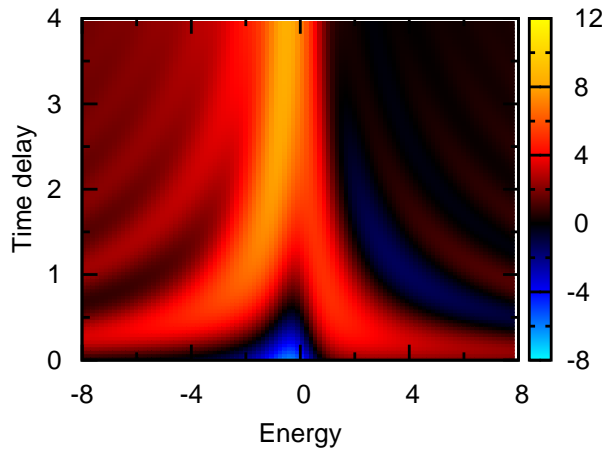}
\caption{Same as Fig.~\ref{fig:shortP} but for the photoabsorption, as described by
Eq.~(\ref{eq:shortS}). When the time delay is 0, the spectrum is close to a vertically
flipped image of the original Fano lineshape.}
\label{fig:shortS}
\end{figure}

In order to demonstrate the influence of laser strength more clearly, the electron and absorption
time-delayed spectra for $A_L=2\pi$ are plotted in Figs.~\ref{fig:shortP} and \ref{fig:shortS},
respectively. As seen in both figures, when $\tau$ is large, very little change is brought by
the laser onto the original Fano lineshape. However, in the $\tau \to 0$ limit,
the electron spectrum flips in the energy dimension around $\epsilon=0$, and the absorption
spectrum displays a strong emission line near $\epsilon=0$.

The significant change of spectral features at $\tau=0$ can be expressed analytically by reducing
Eqs.~(\ref{eq:shortP}) and (\ref{eq:shortS}) to
\begin{align}
P(\epsilon) &= \left|A_X\right|^2 \frac{(-q+\epsilon)^2+4}{1+\epsilon^2}, \label{eq:shortP2}\\
S(\epsilon) &= \left|A_X\right|^2 \left[ -\frac{(q+\epsilon)^2}{1+\epsilon^2}+2 \right]. \label{eq:shortS2}
\end{align}
In the electron spectrum, the mirror image of Fano resonance, indicated by the minus sign of
the $q$ parameter, is added onto a Lorentzian shape. In the absorption spectrum, beside an
additional constant background, the whole resonance function flips upside-down, which means that
the absorption line becomes the emission line. Note that Eqs.~(\ref{eq:shortP2}) and
(\ref{eq:shortS2}) are the results for a full Rabi cycle and represent the extreme changes that
the strong coupling can bring. For non-integer number of cycles, among the two resonances,
some electrons will stay at $|b_2\rangle$ at the end of the laser, and will not be combined with
the photoelectrons in the neighborhood of $|b_1\rangle$.

\end{document}